\documentclass[reprint,superscriptaddress,amsmath,amssymb,aps,pre]{revtex4-2}
\usepackage[utf8]{inputenc}
\usepackage[T1]{fontenc}
\usepackage{graphicx}
\usepackage{dcolumn}
\usepackage{bm}
\usepackage[table]{xcolor}
\usepackage{float}
\usepackage{multirow}
\usepackage{amsfonts}
\usepackage{subcaption}
\usepackage{booktabs}
\usepackage{url}
\usepackage[hidelinks]{hyperref}
\usepackage{orcidlink}
\usepackage{placeins}

\newcommand{\dalpha}{\ensuremath{\Delta\alpha}}
\newcommand{\apeak}{\ensuremath{\alpha_{\mathrm{peak}}}}
\newcommand{\aasym}{\ensuremath{\alpha_{\mathrm{asym}}}}
\newcommand{\vphi}{\ensuremath{v_{\phi}}}
\newcommand{\vx}{\ensuremath{v_{x}}}
\newcommand{\sci}[2]{\ensuremath{#1\times10^{#2}}}

\begin{document}

\title{Multifractality of human upper-limb velocity tracks motor optimization via hand dominance experiment}

\author{Yago Emanoel Ramos \orcidlink{0009-0003-7125-8199}}
\email{yago.emanoel@ufba.br}
\affiliation{Biosystems Laboratory, Instituto de Física, Universidade Federal da Bahia, Salvador, Brazil}
\affiliation{Department of Theoretical Physics, Faculty of Sciences, University of Zaragoza, Zaragoza, Spain}
\affiliation{Institute for Biocomputation and Physics of Complex Systems (BIFI), University of Zaragoza, Zaragoza, Spain}

\author{José Garcia Vivas Miranda \orcidlink{0000-0002-7752-8319}}
\email{vivas@ufba.br}
\affiliation{Biosystems Laboratory, Instituto de Física, Universidade Federal da Bahia, Salvador, Brazil}

\begin{abstract}
Theoretical goal-directed movements follow a unique, stereotyped velocity profile, but continuous movement has no unique optimal solution, so it is unclear whether motor optimization makes continuous movement more stereotyped or more diverse. Previous studies indicated higher information content in the dominant hand, suggesting greater diversity. Because continuous velocity is a composition of overlapping joints movement of varying form, timing and amplitude, comparable to the events that compose spike trains and heartbeat series, we asked whether this diversity appears as a wider multifractal spectrum of velocity. We applied multifractal detrended fluctuation analysis to motion-capture velocity of cursive handwriting from right- and left-handed participants writing with both hands; hand dominance is a natural experiment in which task and geometry are fixed while optimization varies. With the dominant hand, spectra were wider, had a lower dominant exponent and were more positively asymmetric in both groups, whereas hands with the same functional role did not differ between groups. A logistic regression on the three spectral features identified the dominant hand with a cross-validated AUC of up to 0.93. Optimized continuous movement is thus not more stereotyped but more diverse: complex, neither stereotyped nor random, with multifractality that tracks hand dominance.
\end{abstract}

\maketitle

\section{Introduction}

Biologically optimized movement has been investigated in robotics and physiology for decades. Isolated, goal-directed movements are analytically tractable because their boundary conditions are fully specified: under the minimum-jerk criterion, a point-to-point movement that starts and ends at rest, with zero velocity and zero acceleration at both ends, has a unique bell-shaped velocity profile determined only by its displacement $\Delta q$ and duration $t_f$~\cite{Flash1688,hoff_model_1994},
\begin{equation}
\dot{q}(t) = \Delta q \left[ 30 \frac{t^4}{t_f^5} - 60 \frac{t^3}{t_f^4} + 30 \frac{t^2}{t_f^3} \right],
\label{eq:hoff}
\end{equation}
where $q$ is any generalized coordinate (derivation in Appendix~\ref{sec:appendix_hoff}). Whether these principles extend to complex, continuous movements and what ``optimized'' means for them remains an open question~\cite{annurev:/content/journals/10.1146/annurev-control-060117-105206, doi:10.1177/02783649231201196,ModelingHumanSuboptimalControlAReview, annurev:/content/journals/10.1146/annurev-bioeng-102723-020454}.

A bridge between both regimes is Movement Element Decomposition (MED), which proposes that continuous movement is a composition of elemental submovements each bounded by instants of zero velocity, as a classical goal-directed action is~\cite{miranda_complex_2018}. However, linking elements to the minimum-jerk model has consequences. In continuous movement, velocity vanishes at element boundaries, but acceleration generally does not~\cite{doi:10.1177/02783649231201196}. Dropping the two acceleration constraints of  Eq.~\ref{eq:hoff} means that minimizing jerk no longer yields a single velocity profile but a broader family of valid solutions. For continuous movement, therefore, optimality does not single out one velocity profile, and nothing in the minimum-jerk framework requires optimized continuous movement to be stereotyped. 
Intuition from motor learning nonetheless points toward stereotypy: practice tends to reduce variability. Thus, a limb more optimized for a task, as the dominant limb is often assumed to be~\cite{https://doi.org/10.1111/ejn.14649}, might be expected to converge toward a template-like velocity. Previous evidence points the other way. Permutation entropy analysis of the time series of the dynamic exponent $\alpha$ relating element displacement to mean velocity across MED elements ($\bar{v} \propto D^{\alpha}$), showed that this series is more complex in tasks performed with the dominant hand~\cite{RAMOS2025117412}. That result, however, concerns the scaling between elements, not the velocity signal itself. This raises the question addressed here: is the dominant limb, presumably more optimized for the task, also more diverse at the level of the velocity itself, rather than more stereotyped? Answering it requires metrics that quantify how velocity variability is organized across time scales, something minimum-jerk-type models, which prescribe the shape of each element, cannot provide.

Here we argue that the velocity of a continuous movement shares the phenomenology of well-studied physiological event trains. In neural activity, discrete action potentials of variable timing and size summate into macroscopic signals such as local field potentials. In cardiac dynamics, heartbeats with variable inter-beat intervals and amplitudes form the heart rate variability (HRV) series. Likewise, signed velocity along a coordinate is a train of approximately bell-shaped movement elements with alternating polarity, whose timing, duration and amplitude vary (amplitudes scale with displacement~\cite{miranda_complex_2018}), and which overlap and differ in shape because boundary accelerations are not forced to zero. As in the neural and cardiac cases, such signals are rhythmic but not strictly periodic~\cite{doi:10.1073/pnas.012579499, rock_quantifying_2025}. In those systems, multifractality, scaling behavior that varies locally in time rather than being captured by a single exponent, is thought to emerge from the nonlinear interaction of microscale components rather than from noise~\cite{voss_methods_2008,e22030309}. Unlike heartbeats or spikes, movement elements have no known pacemaker or threshold mechanism, so we use the analogy as a phenomenological guide that connects the extensive literature on these signals to movement, not as a claim of shared mechanism. Under this view, the more diverse the element parameters a limb actually uses, within the family permitted by relaxed boundary conditions, the more heterogeneous the local scaling of its velocity series, that is, the wider its multifractal spectrum.

Multifractal analysis~\cite{ivanov_multifractality_1999} quantifies precisely this heterogeneity of local scaling, which makes it the natural framework for the present question. Biological complexity is generally associated with scale-free behavior across many temporal scales rather than with a single parameter regime~\cite{PhysRevLett.89.068102,PhysRevE.71.021906, PhysRevE.78.020901}, and multifractal analysis is widely applied to HRV~\cite{voss_methods_2008,shekatkar_detecting_2017,mangalam_multifractal_2023} and neural signals~\cite{10.1371/journal.pone.0068360,10.3389/fphys.2018.01767, 14z6-hzy3, FRANCA2019612, ramos2026multifractalhumansignalsedge}. Movement also shows functional nonlinear fluctuations~\cite{STERGIOU2011869,GILFRICHE2019102518,schlattmann_angular_2025}, and multifractality has been reported in body-part displacement during object manipulation~\cite{mangalam_multifractal_2020}. We extend this approach to signed velocity, the quantity in which movement elements are defined (Eq.~(1)) and which has been associated with motor optimization~\cite{https://doi.org/10.1111/ejn.14649, RAMOS2025103425,RAMOS2026118081,doi:10.1177/15500594251394773,SILVA2026110099}. The geometry of the multifractal spectrum then summarizes movement complexity: its width ($\dalpha$) measures how diverse the local scaling behaviors are; the position of its peak ($\apeak$) gives the dominant regularity, with lower values indicating rougher, less regular fluctuations; and its asymmetry ($\aasym$) indicates whether large or small fluctuations contribute more to the multifractal structure.

Hand dominance offers a natural experiment to test whether these features reflect motor optimization: the same participant performs the same task, with the same geometry, using a more and a less optimized limb. Previous analyses of this dataset support this association between dominance and motor optimization~\cite{RAMOS2025117412,RAMOS2025103425}. We analyzed cursive handwriting recorded by motion capture in right- and left-handed participants writing with both hands; including both handedness groups allows the functional role of the hand (dominant or non-dominant) to be dissociated from its side. Our primary objective is to test whether the multifractality of upper-limb velocity is consistent across individuals and sensitive to hand dominance, as a signature of the complexity and diversity of optimized movement. We hypothesize that (i)~the velocity of continuous movement is multifractal, with spectra consistent across participants and distinguishable from surrogate data; (ii)~the dominant hand is not more stereotyped but more diverse, with a wider spectrum (larger $\dalpha$), and with $\apeak$ and $\aasym$ also differing systematically from the non-dominant hand; and (iii)~these differences depend on the functional role of the hand rather than on handedness group, so that the dominant hands of right- and left-handers resemble each other, as do their non-dominant hands.

\section{Materials and methods}
\subsection{Experimental data and kinematics}

\begin{figure}
\centering
\includegraphics[width=0.5\textwidth]{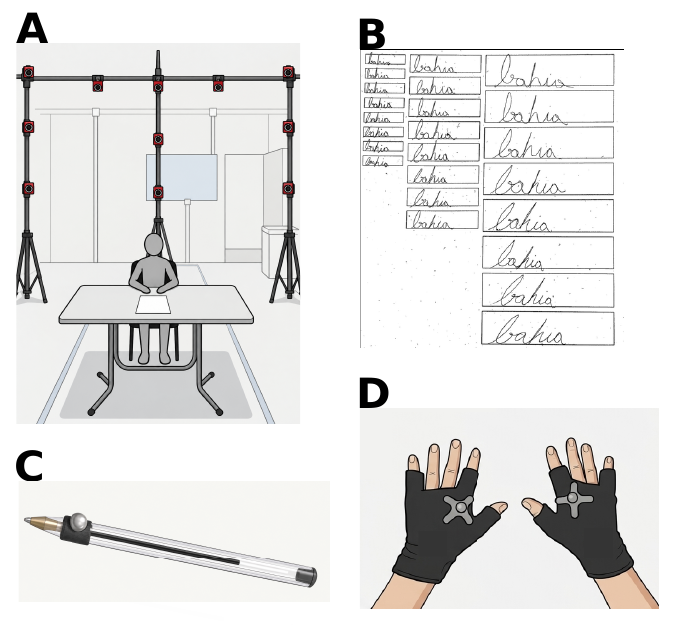}
\caption{\textbf{Experimental setup.} (\textbf{A})~The MOCAP room and experimental arrangement, with the participant seated at the table facing away from the experimenters. (\textbf{B})~The response sheet containing boxes of three sizes that participants were instructed to fill by repeatedly writing the word \textit{bahia} in cursive lowercase; the smallest box measures approximately $2.4\,\text{cm}$ wide and $0.6\,\text{cm}$ tall, with the two larger sizes following the same aspect ratio at $2\times$ and $4\times$ the baseline dimensions, respectively. (\textbf{C})~The pen and its marker, positioned at the point where the fingertips naturally grip the pen, tracking the distal end of the writing instrument. (\textbf{D})~Placement of the metacarpal markers on the base of the index finger of each hand, mounted over the motion capture gloves.}
\label{fig:setup}
\end{figure}

This study uses experimental data collected at the School of Dance of the Federal University of Bahia (Fig.~\ref{fig:setup}A). Data were acquired in a dedicated motion capture (MOCAP) room equipped with a 3D infrared 17-camera system operating at 120~Hz, which tracked the spatial position of reflective markers with a calibrated precision of 1~mm. The system was calibrated daily before the first recording session of each day. In all trials, participants sat on a chair facing a table, wearing gloves fitted with markers positioned on the metacarpal bone of the index finger. Subjects sat with their backs to the room entrance and to the experimenters, minimizing potential visual distractions during the task.

The dataset comprised 70 adult participants (36 self-reported right-handed and 34 left-handed; age $24.71 \pm 7.76$ years, mean $\pm$ SD). Ten participants were excluded (remaining: 28 right-handers and 32 left-handers) because of sensor occlusion by the pen marker or because the glove marker, which was not used in the first collection sessions, was absent. Depending on the individual's writing posture, leaning over the table could obstruct the infrared cameras and produce excessive data gaps in the capture stream. Only participants with at least one minute of continuous artifact-free data were retained; this one-minute window ($N = 7200$ samples) was adopted as the standard analysis length, and all time series were truncated accordingly. Among the 70 recruited participants, no statistically significant differences were found between handedness groups in age (independent-samples $t$-test, $p = 0.547$) or biological sex (46 males and 24 females; chi-square test, $p = 0.671$). The number of observations varies slightly across comparisons and is reported in each table.

The task consisted of repeatedly writing the word \textit{bahia} (the name of both the state and the university where the experiment took place), a word familiar to all participants, in lowercase cursive script. Cursive handwriting was chosen to elicit smoother, more continuous finger movements than print script, whose predominantly straight strokes can introduce abrupt discontinuities or motion artifacts; it is also the culturally dominant writing style in Brazil (Fig.~\ref{fig:setup}B). Subjects performed the task once with each hand, in randomized order, generating four group-task combinations:

\begin{description}
\item[RH$\_$R] right-handers writing with the right hand (dominant)
\item[RH$\_$L] right-handers writing with the left hand (non-dominant)
\item[LH$\_$R] left-handers writing with the right hand (non-dominant)
\item[LH$\_$L] left-handers writing with the left hand (dominant)
\end{description}

Three kinematic representations of the same recording were analyzed: the \textit{pen} condition (position of the pen marker, Fig.~\ref{fig:setup}C), the \textit{hand} condition (position of the metacarpal marker, Fig.~\ref{fig:setup}D), and the \textit{finger} condition, the position of the pen relative to the metacarpal marker. The finger condition removes gross hand displacement and isolates the movement of the fingers holding the pen, since the pen marker lies near the finger-pen contact region; it was used for all statistical analyses. Cartesian velocity components were obtained by finite differences and converted to cylindrical and spherical components (Appendix~\ref{sec:appendix_coordinates}). In the finger condition the metacarpal marker is the origin of these coordinate systems. The azimuthal angle $\phi$ best isolates finger movement relative to the hand, because it describes angular motion in the $xy$-plane with minimal variation in the radial distance between the metacarpal marker and the pen tip. Although the word is written along the $x$-axis, its vertical extent is captured by the $y$-axis, and the angular, smoothly curved strokes of cursive script produce rich $\phi$ dynamics. We therefore report two examples in detail: the tangential velocity $\vphi$ and the Cartesian velocity along the writing direction $\vx$. All other components are shown for an example recording in Fig.~\ref{fig:mfdfa}.

\subsection{Multifractal Detrended Fluctuation Analysis (MFDFA)}

The multifractal properties of the time series were evaluated using Multifractal Detrended Fluctuation Analysis (MFDFA)~\cite{KANTELHARDT200287}. For a time series $x_i$ of length $N$, the algorithm first computes the cumulative centered profile $y(i) = \sum_{k=1}^i (x_k - \bar{x})$, where $\bar{x}$ is the series mean. The profile is divided into $N_s = \lfloor N/s \rfloor$ non-overlapping segments of length $s$. Because $N$ is generally not a multiple of $s$, a short portion at the end of the profile may be discarded; to avoid losing data, the segmentation is repeated starting from the opposite end of the sequence, resulting in $2N_s$ segments.

We used 25 logarithmically spaced scales $s$ between $s_{\min} = \max(16, \lfloor N/1024 \rfloor)$ and $s_{\max} = \lfloor N/4 \rfloor$. For $N = 7200$ at 120~Hz this corresponds to 16-1800 samples (0.13-15~s). Within each segment $\nu = 1, \dots, 2N_s$, a local linear trend ($m=1$), $y_{\mathrm{fit},\nu}$, is fitted and the variance of the detrended profile is computed:
\begin{equation}
F^2(s, \nu) = \frac{1}{s} \sum_{i=1}^s \left[y((\nu-1)s + i) - y_{\mathrm{fit}, \nu}(i)\right]^2 .
\end{equation}
The $q$-th order fluctuation function is obtained by averaging over all segments, as in detrended fluctuation analysis~\cite{PhysRevE.49.1685, REBOREDO20131631}. The moment order $q$ took 1001 linearly spaced values in $[-5, 5]$. For $\vert q \vert \ge 0.05$,
\begin{equation}
F_q(s) = \left\{ \frac{1}{2N_s} \sum_{\nu=1}^{2N_s} \left[F^2(s, \nu)\right]^{q/2} \right\}^{1/q},
\end{equation}
and, to avoid divergence, for $\vert q \vert < 0.05$ the limit $q \to 0$ is approximated by the logarithmic average
\begin{equation}
F_0(s) = \exp \left\{ \frac{1}{4N_s} \sum_{\nu=1}^{2N_s} \ln \left[F^2(s, \nu)\right] \right\}.
\end{equation}
The generalized Hurst exponent $H(q)$ is the slope of the linear regression of $\ln F_q(s)$ on $\ln s$, from $F_q(s) \sim s^{H(q)}$. The mass exponent is $\tau(q) = qH(q) - 1$, and the singularity strength (Hölder exponent) $\alpha$ and singularity spectrum $f(\alpha)$ follow from the Legendre transformation
\begin{equation}
\alpha = \frac{d\tau(q)}{dq}, \qquad f(\alpha) = q\alpha - \tau(q),
\end{equation}
where the derivative is computed numerically with a univariate smoothing spline on $\tau(q)$~\cite{surfaces9010014}.

\begin{figure}
\centering
\includegraphics[width=0.5\textwidth]{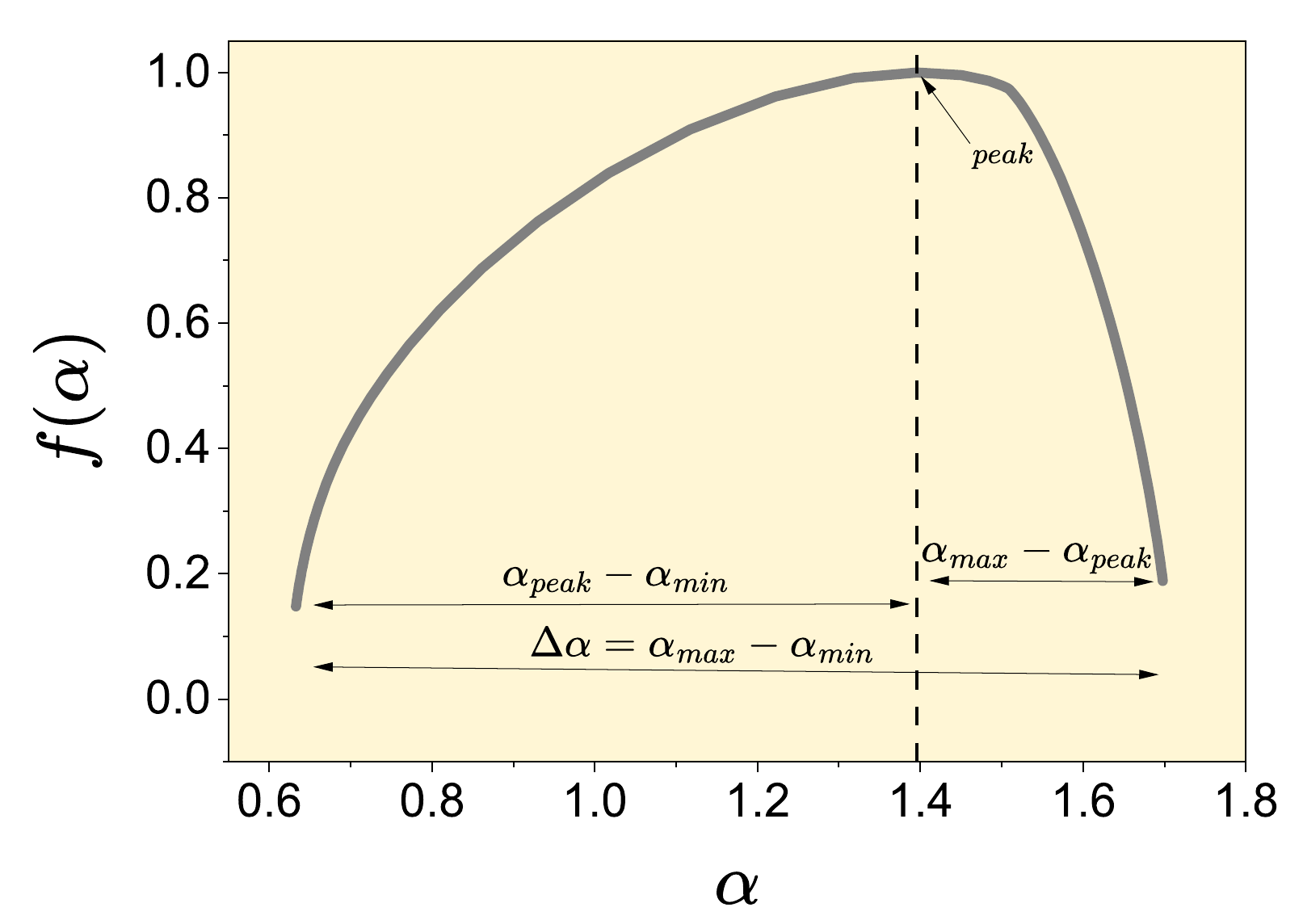}
\caption{\textbf{Multifractal spectrum and the features extracted from it.} The spectrum width $\dalpha = \alpha_{\max}-\alpha_{\min}$, the peak position $\apeak$, and the branch lengths $\apeak-\alpha_{\min}$ and $\alpha_{\max}-\apeak$ used to compute the asymmetry $\aasym$.}
\label{fig:spectrum}
\end{figure}

Three features describe the geometry of the spectrum (Fig.~\ref{fig:spectrum}). The width, $\dalpha = \alpha_{\max} - \alpha_{\min}$, quantifies the degree of multifractality: a wider spectrum indicates a broader distribution of local scaling behaviors. The dominant exponent, $\apeak$, is the value of $\alpha$ at which $f(\alpha)$ is maximal and reflects the global regularity and dominant correlation structure of the series; $\apeak \approx 0.5$ for uncorrelated noise, and larger values indicate smoother, more persistent dynamics. The asymmetry\cite{PhysRevE.91.030902},
\begin{equation}
\aasym = \frac{(\apeak - \alpha_{\min}) - (\alpha_{\max} - \apeak)}{\dalpha},
\end{equation}
is positive when the left branch of the spectrum ($\alpha < \apeak$) is longer than the right branch, and negative otherwise. The left branch is generated by $q>0$, which weights segments with large fluctuations, so a positive asymmetry indicates that the multifractal structure is dominated by the heterogeneity of large fluctuations, and a negative asymmetry that it is dominated by small fluctuations.

\subsection{Statistical comparisons}

Features were compared between the four group-task combinations. Comparisons within a handedness group (RH$\_$R vs.\ RH$\_$L and LH$\_$R vs.\ LH$\_$L) involve the same participants and were tested with the Wilcoxon signed-rank test; comparisons between groups (independent participants) were tested with the Mann-Whitney $U$ test. $p$-values  were Bonferroni-corrected (multiplied by six) within each feature. Effect sizes are the matched-pairs rank-biserial correlation (Wilcoxon) and Cliff's $\delta$ (Mann-Whitney), both positive when the first distribution is larger. Distributions are summarized as mean $\pm$ standard error.

Multifractality was also compared with shuffled surrogates. For each velocity series (finger condition), 10 surrogates were generated by random permutation of their samples, which preserves the amplitude distribution but destroys temporal correlations; the same MFDFA features were computed. Because each participant contributes one series per hand, features were averaged over the two hands of each participant (for surrogates, also over the 10 realizations), so that the unit of analysis is the participant. Original and surrogate values were compared with the Wilcoxon signed-rank test, Bonferroni-corrected over six tests (three features, two velocities).

\subsection{Classification of the dominant hand}

To assess how well the features discriminate dominant from non-dominant hand execution, we ran a Receiver Operating Characteristic (ROC) analysis separately for left- and right-handed participants. First, for each single feature ($\dalpha$, $\apeak$, $\aasym$), the true positive rate (sensitivity) was plotted against the false positive rate ($1-$specificity) across thresholds and the Area Under the Curve (AUC) was computed; features with an initial AUC $<0.5$ were inverted to report their absolute discriminatory power.

Second, the three features were combined in a logistic regression, with $x_1 = \dalpha$, $x_2 = \apeak$ and $x_3 = \aasym$. The log-odds of a trial belonging to the positive class $y=1$ (dominant-hand execution) is a linear combination of the features~\cite{bewick_statistics_2005},
\begin{equation}
z = \beta_0 + \beta_1 \dalpha + \beta_2 \apeak + \beta_3 \aasym,
\end{equation}
where $\beta_0$ is the intercept and $\beta_1, \beta_2, \beta_3$ are the learned coefficients, and the predicted probability is given by the logistic function,
\begin{equation}
p = P(y=1 | \mathbf{x}) = \frac{1}{1 + e^{-z}}.
\end{equation}
The combined model was evaluated by leave-one-subject-out (LOSO) cross-validation: for each participant, the model was fitted on the trials of all other participants of the same group, and the predicted probabilities for the two trials of the held-out participant were stored. The ROC curve and AUC of the combined model were computed from the pooled held-out probabilities, which gives a consistent estimate of out-of-sample performance in small samples. Single-feature AUCs involve no fitting other than the sign of the effect.

\section{Results}

\subsection{Velocity of continuous writing is multifractal}

To test whether the spectra depend on the temporal organization of the velocity rather than on its amplitude distribution alone, we compared each series with shuffled surrogates (Table~\ref{tab:shuffle_tests}). For both $v_\phi$ and $v_x$, all three features differed from those of the surrogates. Because each participant contributes two tasks, the comparison used one value per participant (averaged over the two hands), and the largest corrected $p$-value was $1.17\times10^{-5}$. The original series had wider spectra (median $\Delta\alpha$ 0.84 vs.\ 0.20 for $v_\phi$ and 0.94 vs.\ 0.31 for $v_x$) and larger $\alpha_{\mathrm{peak}}$ (0.70 vs.\ 0.52 and 0.76 vs.\ 0.53), whereas the surrogates had $\alpha_{\mathrm{peak}}$ close to 0.5, the value expected for uncorrelated noise. The asymmetry also differed from that of the surrogates, but in opposite directions for the two velocities (larger for $v_\phi$, smaller for $v_x$), so its sign is a property of each coordinate and not a general feature of the movement. These results support the first hypothesis: the multifractal spectra of handwriting velocity arise from the temporal organization of the signal.

\begin{table*}[htbp]
\centering
\begin{tabular}{l>{\columncolor{gray!20}}lr>{\columncolor{gray!20}}cc>{\columncolor{gray!20}}rr}
\toprule
Velocity & Metric & $n$ & Original & Shuffled & $p$ & $p_{\mathrm{Bonf.}}$ \\
\midrule
$v_{\phi}$ & $\Delta\alpha$ & 60 & 0.840 [0.745, 0.968] & 0.196 [0.151, 0.237] & 1.11e-11 & 6.67e-11 \\
 & $\alpha$-peak & 60 & 0.703 [0.634, 0.758] & 0.515 [0.511, 0.525] & 1.42e-11 & 8.55e-11 \\
 & $\alpha$-asymmetry & 60 & 0.204 [0.056, 0.319] & 0.043 [-0.056, 0.109] & 1.22e-6 & 7.33e-6 \\
\midrule
$v_{x}$ & $\Delta\alpha$ & 60 & 0.939 [0.844, 1.038] & 0.314 [0.273, 0.358] & 1.11e-11 & 6.67e-11 \\
 & $\alpha$-peak & 60 & 0.763 [0.714, 0.823] & 0.531 [0.523, 0.541] & 1.11e-11 & 6.67e-11 \\
 & $\alpha$-asymmetry & 60 & -0.039 [-0.129, 0.027] & 0.082 [0.004, 0.154] & 1.95e-6 & 1.17e-5 \\
\bottomrule
\end{tabular}
\caption{Multifractal metrics of the velocity series versus 10 shuffled surrogates (all conditions pooled; one value per subject, averaged over tasks; paired by subject). Values: median [Q1, Q3]. $p$: Wilcoxon signed-rank test. $p_{\mathrm{Bonf.}}$: Bonferroni-corrected over the 6 tests.}
\label{tab:shuffle_tests}
\end{table*}

\begin{figure}
\centering
\includegraphics[width=0.5\textwidth]{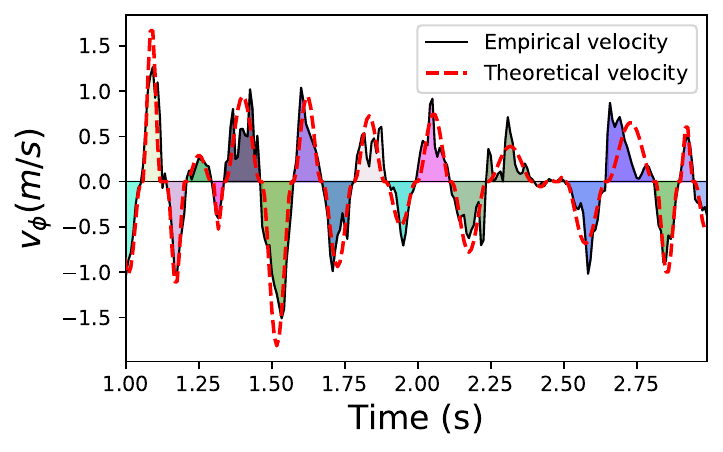}
\caption{\textbf{Azimuthal tangential velocity $\vphi(t)$.} Theoretical velocity (dashed red line) according to the Hoff velocity function adjusted with duration and displacement for each movement element. Colors of elements were assigned randomly in RGB space solely to visually distinguish their boundaries.}
\label{fig:phi}
\end{figure}

Fig.~\ref{fig:phi} shows a short excerpt of $\vphi$ during writing. Some movement elements closely follow the theoretical Hoff profile fitted to their displacement $\Delta q$ and duration $t_f$ (Eq.~\ref{eq:hoff}), but this template does not describe the full sample of continuous movement: element amplitudes and durations vary, and elements overlap with smaller fluctuations. To test whether this variability has a multifractal organization across time scales, we applied MFDFA to each velocity component of several kinematic representations, analyzing each coordinate independently.

\begin{figure*}
\centering
\includegraphics[width=0.75\textwidth]{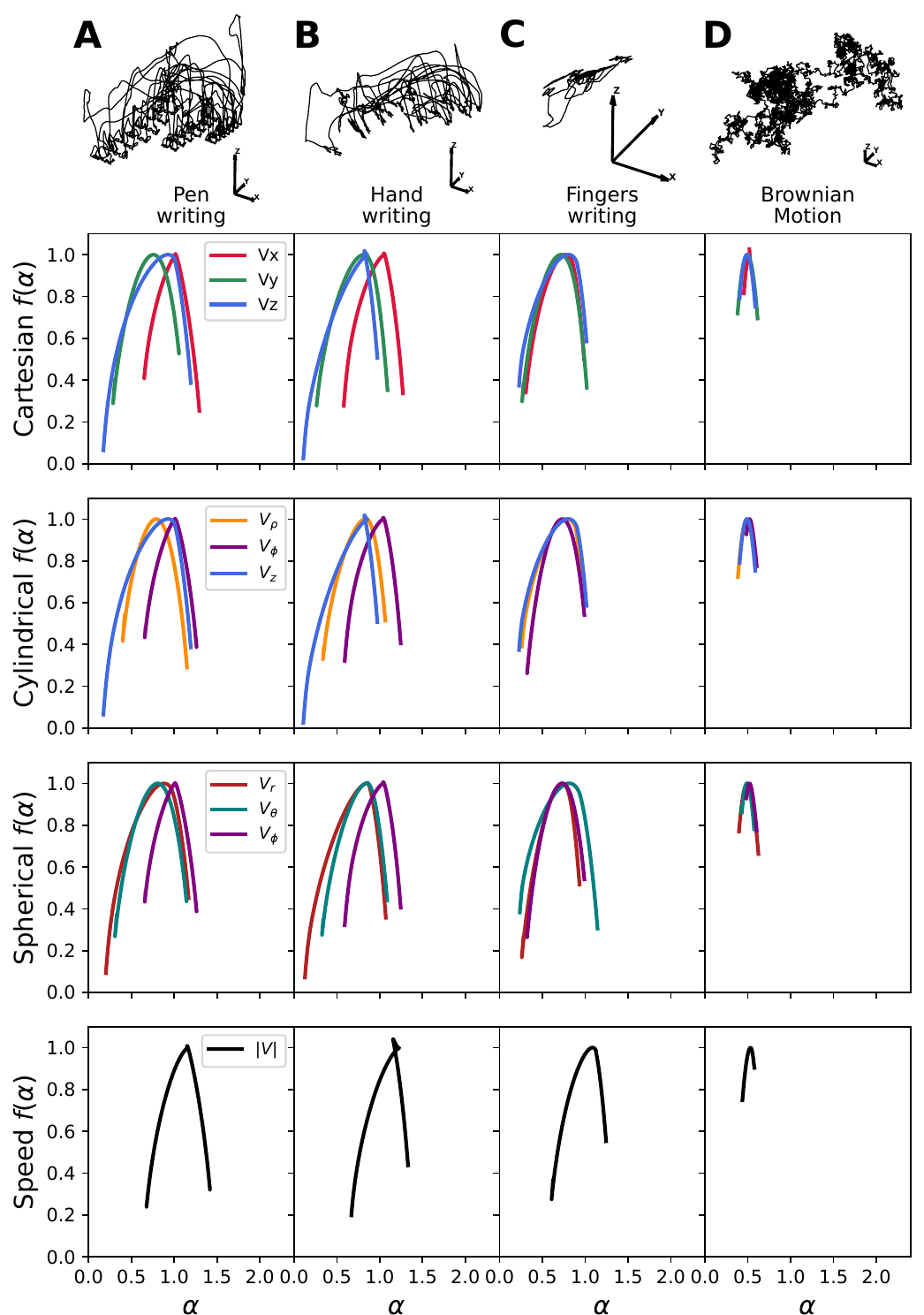}
\caption{\textbf{Examples of MFDFA spectra across kinematic representations, coordinate systems and scalar speed for one subject.} Multifractal spectra of the same handwriting recording computed from (\textbf{A})~the pen marker, (\textbf{B})~the hand (glove) marker, and (\textbf{C})~the position of the pen relative to the hand marker (finger movement), and (\textbf{D})~for simulated Brownian motion. Each column presents the spectra for one condition, whose three-dimensional trajectory is displayed above the corresponding plots; rows show Cartesian, cylindrical and spherical components and the scalar speed $|V|$. Reference axes with straight lines in trajectories represent 3~cm in each Cartesian direction at the scale of the figure.}
\label{fig:mfdfa}
\end{figure*}

The analyzed velocities yielded multifractal spectra with $\dalpha$ exceeding 1 in several recordings (Fig.~\ref{fig:mfdfa}), and the spectra depended on the kinematic representation and on the coordinate.

The pen marker (Fig.~\ref{fig:mfdfa}A) reflects the whole kinematic chain: besides the strokes that form letters, it carries task-related but functionally distinct proximal motion, such as trunk and arm transport along the line and postural stabilization, which can broaden the spectrum independently of movement-element diversity. These movements are better isolated by the hand marker (Fig.~\ref{fig:mfdfa}B). Both the pen and the hand marker show an asymmetry between the coordinates aligned with the writing direction ($x$ and $\phi$) and the remaining ones: $\apeak$ of $x$ and $\phi$ is displaced to the right, which indicates more regular dynamics along the writing axis and its associated rotations in the plane. A corner or fish-tail effect is also visible in some of the spectra of Fig.~\ref{fig:mfdfa}B, a feature frequently associated with spectra produced by isolated singularities rather than by genuine multifractal organization~\cite{oswiecimka_wavelet-based_2020}, which calls for caution when interpreting absolute-marker spectra. In contrast, the movement of the pen relative to the hand (Fig.~\ref{fig:mfdfa}C), which isolates the fine motor task, is more consistent: the spectra of the different coordinates are similar, suggesting that separating finger movement from the hand makes multifractal features more uniform across coordinates. Whether this holds for movements that are less restricted to a plane should be addressed in future work. Scalar speed, which is non-negative and therefore insensitive to reversals in movement direction, tends to peak at larger $\alpha$ than the signed components (bottom row of Fig.~\ref{fig:mfdfa}), i.e.\ it is more regular. Because of its fine motor task isolation, all statistical analyses below use the isolated finger movement condition.

For visual reference, Fig.~\ref{fig:mfdfa}D shows the spectra of simulated Brownian motion (Appendix~\ref{sec:appendix_brownian}), whose velocity is approximately white noise; they are narrow and centered near $\alpha = 0.5$, similar to $\alpha$-peak values found in shuffled velocities (Table~\ref{tab:shuffle_tests}). The simulation is not used in the statistical analyses, which focus on hand dominance.

\FloatBarrier

\subsection{Multifractal features depend on hand dominance}

\begin{figure}
\centering
\includegraphics[width=0.5\textwidth]{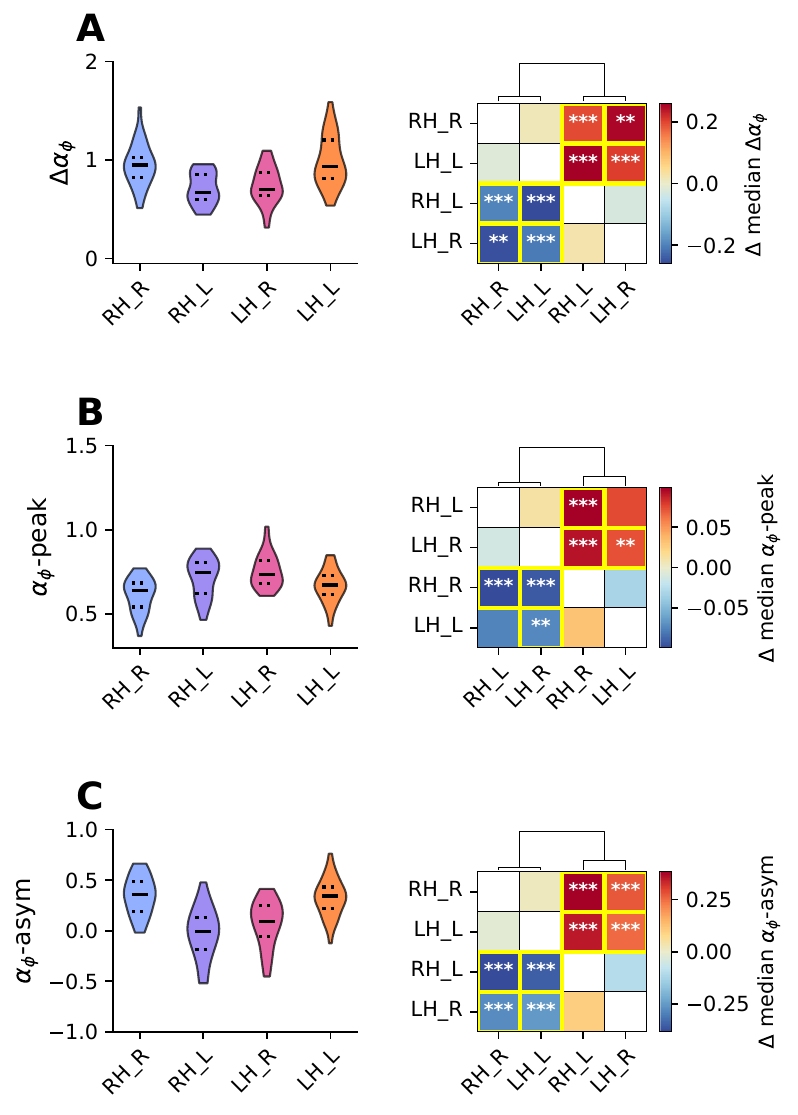}
\caption{\textbf{Distributions and dissimilarities of multifractal features of the azimuthal tangential velocity $\vphi$.} Violin plots (left column) show the distributions across group-task combinations, with the corresponding dissimilarity matrices (right column) ordered by dendrograms. Heatmap colors represent the difference in medians ($\Delta$ median) between conditions. Pairwise comparisons are Bonferroni-corrected, using the tests described in Table~\ref{tab:phi}; significant differences are marked with asterisks (*$p<0.05$, **$p<0.01$, ***$p<0.001$). Features: (\textbf{A})~$\dalpha$; (\textbf{B})~$\apeak$; (\textbf{C})~$\aasym$.}
\label{fig:dist_phi}
\end{figure}

\begin{figure}
\centering
\includegraphics[width=0.5\textwidth]{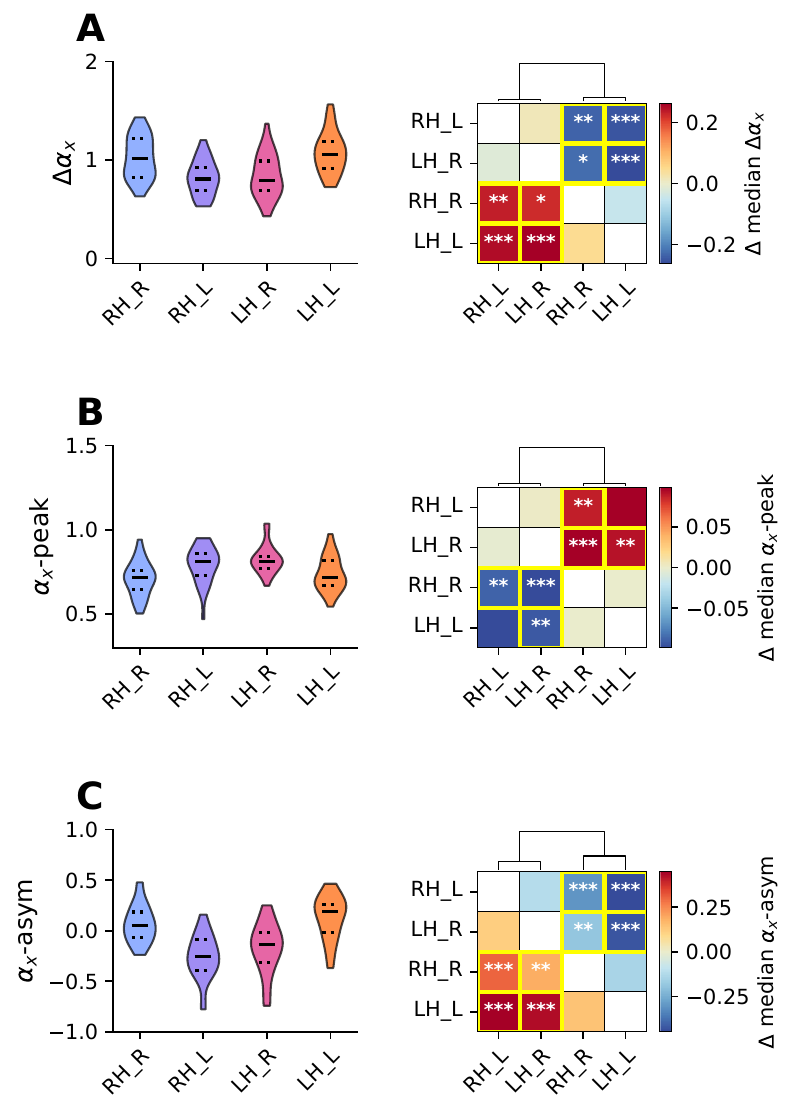}
\caption{\textbf{Distributions and dissimilarities of multifractal features of the Cartesian velocity along the writing direction $\vx$.} Layout, tests and significance marks as in Fig.~\ref{fig:dist_phi}, with statistics in Table~\ref{tab:x}. Features: (\textbf{A})~$\dalpha$; (\textbf{B})~$\apeak$; (\textbf{C})~$\aasym$.}
\label{fig:dist_x}
\end{figure}

The dependence of the spectrum on the coordinate shows that multifractal features are sensitive to the structure of the movement. To test whether they are also sensitive to the degree of motor optimization in movements that share the same geometry, we compared handwriting of right- and left-handed participants with both hands (Figs.~\ref{fig:dist_phi} and~\ref{fig:dist_x}; Tables~\ref{tab:phi} and~\ref{tab:x}), using $\vphi$ as the main example and $\vx$ as a second one.

For $\vphi$, the spectrum was wider with the dominant hand (Fig.~\ref{fig:dist_phi}A; Table~\ref{tab:phi}). The difference between dominant and non-dominant hand was significant in all four comparisons that cross the dominant/non-dominant role (RH$\_$R vs.\ RH$\_$L, RH$\_$R vs.\ LH$\_$R, RH$\_$L vs.\ LH$\_$L, LH$\_$R vs.\ LH$\_$L; $p_{\mathrm{Bonf.}} \le \sci{2.6}{-3}$), with large effect sizes in the within-subject comparisons ($|r| = 0.84$-$0.92$). The asymmetry $\aasym$ was more positive with the dominant hand, with the same pattern of significance ($p_{\mathrm{Bonf.}} \le \sci{2.4}{-4}$ in all four comparisons; Fig.~\ref{fig:dist_phi}C). Finally, $\apeak$ was lower with the dominant hand (Fig.~\ref{fig:dist_phi}B), significantly so in three of the four comparisons; the exception, RH$\_$L vs.\ LH$\_$L, shows the same direction (mean 0.715 vs.\ 0.668) but does not show significance ($p_{\mathrm{Bonf.}} = 0.48$). Comparisons between hands of equivalent functional role, dominant hands (RH$\_$R vs.\ LH$\_$L) or non-dominant hands (RH$\_$L vs.\ LH$\_$R) of the two handedness groups, were non-significant even before correction for all three features (smallest uncorrected $p = 0.072$).

The velocity $\vx$ gave analogous results (Fig.~\ref{fig:dist_x}; Table~\ref{tab:x}): larger $\dalpha$ and more positive $\aasym$ with the dominant hand in all four cross-role comparisons, lower $\apeak$ in three of four (RH$\_$L vs.\ LH$\_$L: $p_{\mathrm{Bonf.}} = 0.15$), and no differences between hands of equivalent functional role (smallest uncorrected $p = 0.10$). The values of $\apeak$ were higher for $\vx$ (0.71-0.82) than for $\vphi$ (0.61-0.75), in line with the more regular dynamics along the writing axis seen in Fig.~\ref{fig:mfdfa}. In both coordinates, and for both handedness groups, the dominant hand therefore has a wider, more positively asymmetric spectrum with a lower dominant exponent, and these signatures follow the functional role of the hand rather than the handedness group.

\begin{table*}[htbp]
\centering

\begin{tabular}{l>{\columncolor{gray!20}}ll>{\columncolor{gray!20}}ll>{\columncolor{gray!20}}rr>{\columncolor{gray!20}}rr}
\toprule
Comparison & Test & $n$ & Distribution 1 & Distribution 2 & $W$ / $U$ & $p$ & $p_{\mathrm{Bonf.}}$ & Effect \\
\midrule
\multicolumn{9}{l}{\textbf{$\Delta \alpha_{\phi}$}} \\
\midrule
RH$\_$R vs. RH$\_$L & WC & 27 & $0.929 \pm 0.036$  & $0.706 \pm 0.030$  & 16.0 & 2.52e-6 & 1.51e-5 & 0.92 \\
RH$\_$R vs. LH$\_$R & MW & 28/32 & $0.951 \pm 0.041$ & $0.742 \pm 0.033$  & 686.0 & 4.33e-4 & 2.60e-3 & 0.53 \\
RH$\_$R vs. LH$\_$L & MW & 28/31 & $0.951 \pm 0.041$ & $0.982 \pm 0.046$ & 422.0 & 0.861 & 1 & -0.03 \\
RH$\_$L vs. LH$\_$R & MW & 28/32 & $0.715 \pm 0.030$  & $0.742 \pm 0.033$  & 401.0 & 0.491 & 1 & -0.10 \\
RH$\_$L vs. LH$\_$L & MW & 28/31 & $0.715 \pm 0.030$  & $0.982 \pm 0.046$  & 173.0 & 7.68e-5 & 4.61e-4 & -0.60 \\
LH$\_$R vs. LH$\_$L & WC & 31 & $0.732 \pm 0.032$  & $0.982 \pm 0.046$  & 39.0 & 7.01e-6 & 4.20e-5 & -0.84 \\
\midrule
\multicolumn{9}{l}{\textbf{$\alpha_{\phi}$-peak}} \\
\midrule
RH$\_$R vs. RH$\_$L & WC & 27 & $0.608 \pm 0.019$  & $0.713 \pm 0.022$  & 28.0 & 2.21e-5 & 1.33e-4 & -0.85 \\
RH$\_$R vs. LH$\_$R & MW & 28/32 & $0.614 \pm 0.019$ & $0.753 \pm 0.017$ & 148.0 & 9.09e-6 & 5.45e-5 & -0.67 \\
RH$\_$R vs. LH$\_$L & MW & 28/31 & $0.614 \pm 0.019$ & $0.668 \pm 0.017$  & 315.0 & 0.0721 & 0.432 & -0.27 \\
RH$\_$L vs. LH$\_$R & MW & 28/32 & $0.715 \pm 0.022$  & $0.753 \pm 0.017$  & 395.0 & 0.437 & 1 & -0.12 \\
RH$\_$L vs. LH$\_$L & MW & 28/31 & $0.715 \pm 0.022$ & $0.668 \pm 0.017$  & 550.0 & 0.0796 & 0.477 & 0.27 \\
LH$\_$R vs. LH$\_$L & WC & 31 & $0.749 \pm 0.017$ & $0.668 \pm 0.017$  & 86.0 & 9.82e-4 & 5.89e-3 & 0.65 \\
\midrule
\multicolumn{9}{l}{\textbf{$\alpha_{\phi}$-asym}} \\
\midrule
RH$\_$R vs. RH$\_$L & WC & 27 & $0.351 \pm 0.037$ & $-0.018 \pm 0.049$ & 11.0 & 8.20e-7 & 4.92e-6 & 0.94 \\
RH$\_$R vs. LH$\_$R & MW & 28/32 & $0.344 \pm 0.036$ & $0.072 \pm 0.041$  & 726.0 & 3.93e-5 & 2.36e-4 & 0.62 \\
RH$\_$R vs. LH$\_$L & MW & 28/31 & $0.344 \pm 0.036$  & $0.326 \pm 0.034$ & 464.0 & 0.654 & 1 & 0.07 \\
RH$\_$L vs. LH$\_$R & MW & 28/32 & $-0.024 \pm 0.048$  & $0.072 \pm 0.041$  & 337.0 & 0.102 & 0.609 & -0.25 \\
RH$\_$L vs. LH$\_$L & MW & 28/31 & $-0.024 \pm 0.048$ & $0.326 \pm 0.034$  & 107.0 & 7.19e-7 & 4.32e-6 & -0.75 \\
LH$\_$R vs. LH$\_$L & WC & 31 & $0.081 \pm 0.042$ & $0.326 \pm 0.034$  & 52.0 & 3.57e-5 & 2.14e-4 & -0.79 \\
\bottomrule
\end{tabular}
\caption{Distribution (mean $\pm$ SE) comparisons between conditions multifractal features in $v_{\phi}$. WC: Wilcoxon signed-rank test (paired subjects, statistic $W$); MW: Mann-Whitney $U$ test (independent groups, statistic $U$). $p_{\mathrm{Bonf.}}$: Bonferroni-corrected $p$-value within each metric.}
\label{tab:phi}
\end{table*}

\begin{table*}[htbp]
\centering
\begin{tabular}{l>{\columncolor{gray!20}}ll>{\columncolor{gray!20}}ll>{\columncolor{gray!20}}rr>{\columncolor{gray!20}}rr}
\toprule
Comparison & Test & $n$ & Distribution 1 & Distribution 2 & $W$ / $U$ & $p$ & $p_{\mathrm{Bonf.}}$ & Effect \\
\midrule
\multicolumn{9}{l}{\textbf{$\Delta \alpha_{x}$}} \\
\midrule
RH$\_$R vs. RH$\_$L & WC & 27 & $1.009 \pm 0.039$  & $0.809 \pm 0.033$  & 59.0 & 1.13e-3 & 6.79e-3 & 0.69 \\
RH$\_$R vs. LH$\_$R & MW & 28/32 & $1.024 \pm 0.040$  & $0.839 \pm 0.038$ & 655.0 & 2.21e-3 & 0.0133 & 0.46 \\
RH$\_$R vs. LH$\_$L & MW & 28/31 & $1.024 \pm 0.040$  & $1.077 \pm 0.038$ & 378.0 & 0.4 & 1 & -0.13 \\
RH$\_$L vs. LH$\_$R & MW & 28/32 & $0.819 \pm 0.033$  & $0.839 \pm 0.038$ & 432.0 & 0.818 & 1 & -0.04 \\
RH$\_$L vs. LH$\_$L & MW & 28/31 & $0.819 \pm 0.033$ & $1.077 \pm 0.038$  & 153.0 & 2.06e-5 & 1.24e-4 & -0.65 \\
LH$\_$R vs. LH$\_$L & WC & 31 & $0.822 \pm 0.036$  & $1.077 \pm 0.038$ & 44.0 & 1.35e-5 & 8.11e-5 & -0.82 \\
\midrule
\multicolumn{9}{l}{\textbf{$\alpha_{x}$-peak}} \\
\midrule
RH$\_$R vs. RH$\_$L & WC & 27 & $0.710 \pm 0.020$  & $0.788 \pm 0.021$  & 59.0 & 1.13e-3 & 6.79e-3 & -0.69 \\
RH$\_$R vs. LH$\_$R & MW & 28/32 & $0.708 \pm 0.019$  & $0.826 \pm 0.015$ & 154.0 & 1.37e-5 & 8.21e-5 & -0.66 \\
RH$\_$R vs. LH$\_$L & MW & 28/31 & $0.708 \pm 0.019$ & $0.737 \pm 0.018$  & 383.0 & 0.443 & 1 & -0.12 \\
RH$\_$L vs. LH$\_$R & MW & 28/32 & $0.793 \pm 0.021$  & $0.826 \pm 0.015$  & 405.0 & 0.529 & 1 & -0.10 \\
RH$\_$L vs. LH$\_$L & MW & 28/31 & $0.793 \pm 0.021$  & $0.737 \pm 0.018$  & 582.0 & 0.0252 & 0.151 & 0.34 \\
LH$\_$R vs. LH$\_$L & WC & 31 & $0.824 \pm 0.015$ & $0.737 \pm 0.018$  & 69.0 & 2.15e-4 & 1.29e-3 & 0.72 \\
\midrule
\multicolumn{9}{l}{\textbf{$\alpha_{x}$-asym}} \\
\midrule
RH$\_$R vs. RH$\_$L & WC & 27 & $0.068 \pm 0.036$  & $-0.255 \pm 0.042$ & 22.0 & 7.99e-6 & 4.79e-5 & 0.88 \\
RH$\_$R vs. LH$\_$R & MW & 28/32 & $0.070 \pm 0.035$ & $-0.176 \pm 0.045$ & 700.0 & 1.94e-4 & 1.16e-3 & 0.56 \\
RH$\_$R vs. LH$\_$L & MW & 28/31 & $0.070 \pm 0.035$  & $0.126 \pm 0.039$ & 345.0 & 0.179 & 1 & -0.21 \\
RH$\_$L vs. LH$\_$R & MW & 28/32 & $-0.256 \pm 0.041$  & $-0.176 \pm 0.045$  & 337.0 & 0.102 & 0.609 & -0.25 \\
RH$\_$L vs. LH$\_$L & MW & 28/31 & $-0.256 \pm 0.041$ & $0.126 \pm 0.039$ & 93.0 & 2.36e-7 & 1.42e-6 & -0.79 \\
LH$\_$R vs. LH$\_$L & WC & 31 & $-0.178 \pm 0.046$ & $0.126 \pm 0.039$ & 48.0 & 2.22e-5 & 1.33e-4 & -0.81 \\
\bottomrule
\end{tabular}
\caption{Distribution (mean $\pm$ SE) comparisons between conditions multifractal features in $v_x$. WC: Wilcoxon signed-rank test (paired subjects, statistic $W$); MW: Mann-Whitney $U$ test (independent groups, statistic $U$). $p_{\mathrm{Bonf.}}$: Bonferroni-corrected $p$-value within each metric.}
\label{tab:x}
\end{table*}

\FloatBarrier
\subsection{Spectral features classify the dominant hand}

\begin{figure*}
\centering
\includegraphics[width=0.8\textwidth]{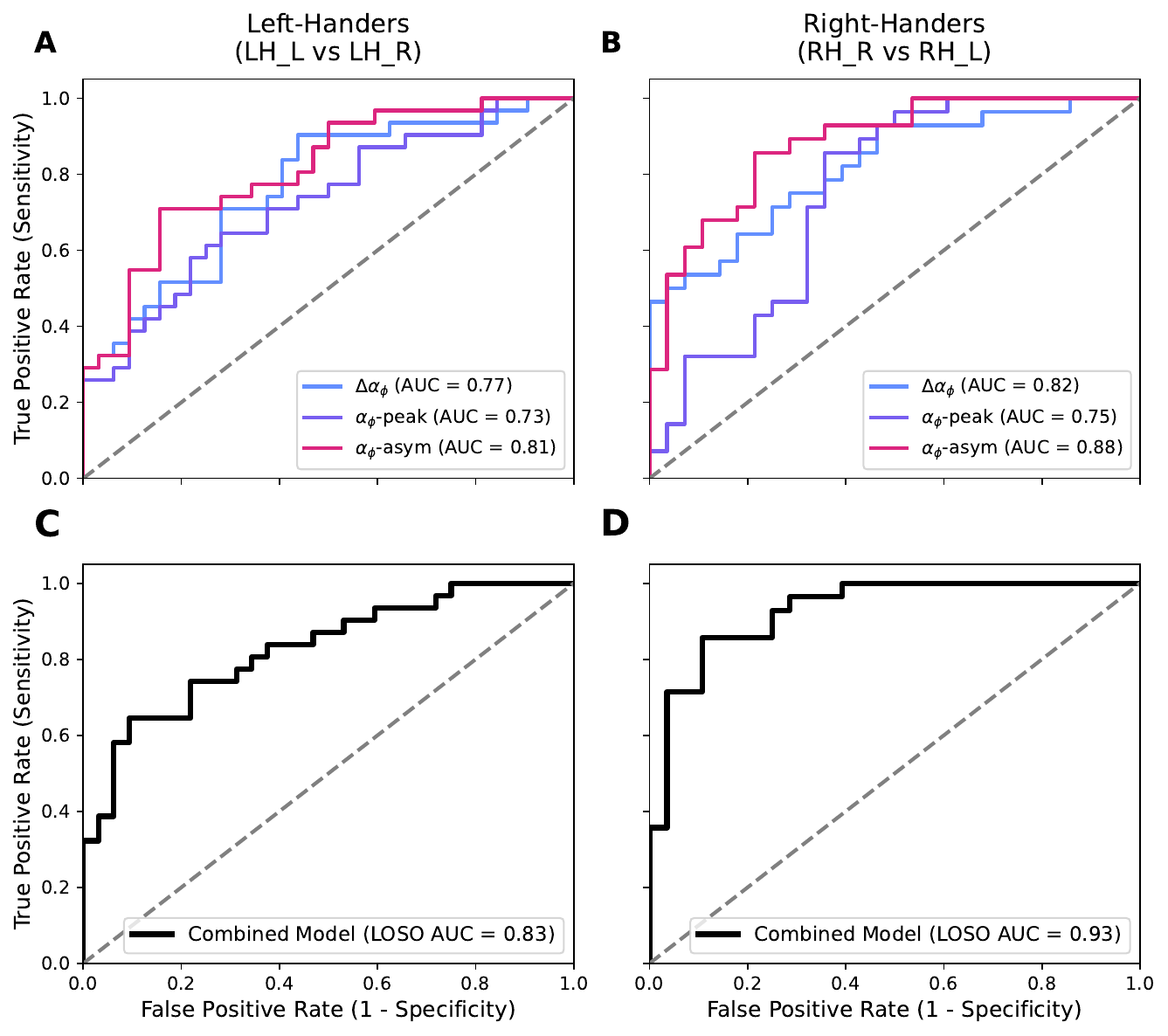}
\caption{\textbf{ROC curves for dominant vs.\ non-dominant hand classification using $\vphi$.} ROC curves are shown separately for left-handers (\textbf{A},\textbf{C}) and right-handers (\textbf{B},\textbf{D}), classifying dominant-hand trials (LH$\_$L and RH$\_$R, respectively) against non-dominant-hand trials (LH$\_$R and RH$\_$L). (\textbf{A},\textbf{B})~Individual ROC curves for each multifractal feature: $\dalpha$ (light blue), $\apeak$ (dark blue) and $\aasym$ (pink). (\textbf{C},\textbf{D})~ROC curve of the logistic regression combining the three features, obtained by leave-one-subject-out (LOSO) cross-validation. The dashed diagonal line is chance level ($\text{AUC}=0.50$). AUC values are given in the legend of each panel.}
\label{fig:roc_phi}
\end{figure*}

\begin{figure*}
\centering
\includegraphics[width=0.8\textwidth]{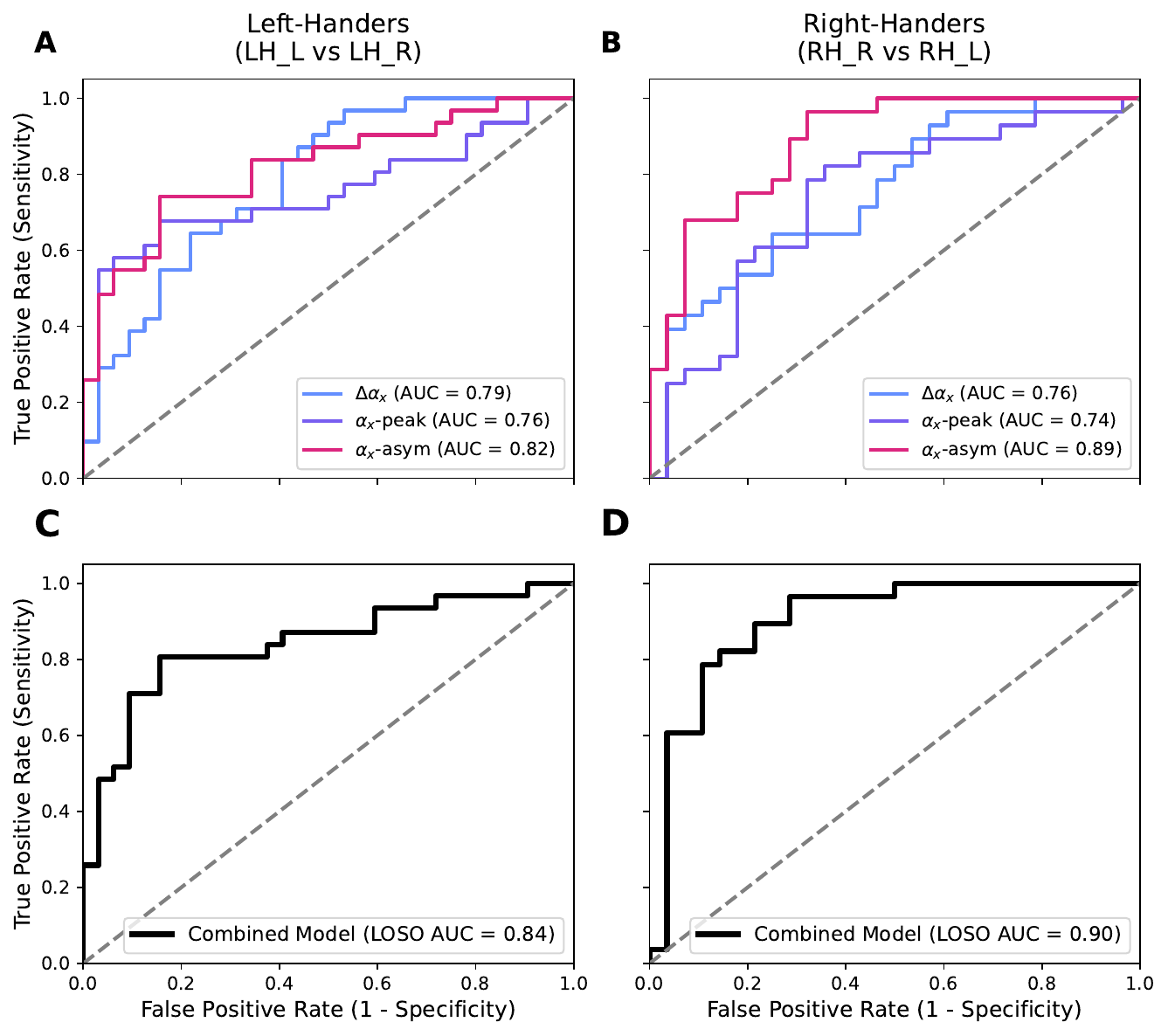}
\caption{\textbf{ROC curves for dominant vs.\ non-dominant hand classification using $\vx$.} Layout as in Fig.~\ref{fig:roc_phi}, with features computed from $\vx$.}
\label{fig:roc_x}
\end{figure*}

As a complementary measure of discriminability, we computed ROC curves for $\vphi$ (Fig.~\ref{fig:roc_phi}) and $\vx$ (Fig.~\ref{fig:roc_x}). The spectral asymmetry gave the highest single-feature AUC in both handedness groups and both coordinates: for $\vphi$, 0.81 in left-handers and 0.88 in right-handers; for $\vx$, 0.82 and 0.89. Combining the three features in a logistic regression, evaluated by leave-one-subject-out cross-validation, gave an AUC of 0.83 (left-handers) and 0.93 (right-handers) for $\vphi$, and 0.84 and 0.90 for $\vx$. The combination therefore improved on the asymmetry alone only modestly, except for $\vphi$ in right-handers (0.88 to 0.93). Thus, the multifractal features discriminate dominant from non-dominant hand trials in both groups, with higher accuracy in right-handers.

\FloatBarrier
\section{Discussion}

We find that the velocity of continuous handwriting has well-defined multifractal spectra (different from random shuffles) and that three geometric features of the spectrum depend on the hand used. For both $\vphi$ and $\vx$, spectra were wider, had a lower dominant exponent and were more positively asymmetric with the dominant hand, in right- and left-handers alike, while hands with the same functional role in the two groups were statistically indistinguishable. The signature therefore follows functional dominance, as expected if it reflects task-specific motor optimization, rather than membership of a handedness group.

These results are consistent with the picture outlined in the Introduction. Relaxing the boundary accelerations of movement elements admits a family of valid velocity profiles, so an optimized limb need not converge on a single template, and the higher information content of the exponent series in the dominant hand~\cite{RAMOS2025117412} points the same way. Our results answer the question raised in the Introduction: the dominant, more optimized hand is not more stereotyped but more diverse. A wider spectrum means a broader range of local scaling behaviors, i.e.\ a more diverse velocity structure. The lower $\apeak$ of the dominant hand, closer to the value 0.5 of uncorrelated noise, means a less smooth, more irregular dominant exponent. Optimization, in this reading, coincides with greater diversity and irregularity rather than with greater regularity. The positive asymmetry of the dominant hand indicates that its multifractal structure is dominated by the heterogeneity of large fluctuations, whereas the non-dominant hand has a nearly symmetric or slightly negative asymmetry. In the event-train picture, this would point to heterogeneity among the largest velocity elements. Hand dominance serves here as the optimization condition, an association established in previous analyses of this dataset~\cite{RAMOS2025117412,RAMOS2025103425}.

The classification results show that these features are informative at the level of a single one-minute recording. A logistic regression on the three features identified the dominant hand with a cross-validated AUC of up to 0.93 in right-handers from raw velocity signals alone. The lower accuracy in left-handers is consistent with the greater variability of motor lateralization reported for left-handers~\cite{RAMOS2025103425}.

The results also bear on comparisons with neural data from the same participants. Previously, differences in kinematic or electroencephalographic (EEG) signals between right- and left-handers were detectable only with relative measures~\cite{RAMOS2025103425,RAMOS2025117412}, which suppress inter-individual variability, until approaches exploiting the directional properties of information exchange in EEG allowed direct group comparisons, indicating that the nature of the measure determines the capacity for group discrimination~\cite{ramos2026corticalinformationtransferreveals}. In those EEG results, differences were associated with the hand being used irrespective of handedness, whereas the multifractal features of movement reported here are associated with hand dominance. This suggests that, although both EEG-based and movement-based measures capture paired differences between tasks performed with the two hands, the EEG differences originate from anatomical lateralization, whereas kinematic signals reflect functional dominance.

Finally, our findings are consistent with the view that velocity fluctuations in human movement are not merely deviations from a theoretical template but carry task-related structure. Human movement is neither stereotyped nor random: its velocity combines movement elements of varying form, timing and amplitude with scale-dependent variability, a hallmark of complex systems shared with cardiac and neural signals.

\section{Limitations and further directions}

A primary limitation is the absence of an anatomically isolated angular coordinate for the $\phi$ axis. Fully isolating the movement of interest would require a rotation matrix anchored to the hand's own reference frame, i.e.\ a glove instrumented with at least three co-planar markers that define a local coordinate system tracking hand rotations. This restricts the present approach to movements largely confined to a planar surface (here, the table), where oscillations between the hand and the pen tip occur approximately along the $\phi$ direction. Future work will extend the setup to arbitrary three-dimensional movements.

\section{Conclusion}

Optimization of continuous movement does not lead to a stereotyped velocity: the dominant, more optimized hand shows more diverse velocity dynamics. The multifractal spectrum of velocity depends on the kinematic representation and is consistently modulated by hand dominance: with the dominant hand, spectra are wider, have a lower dominant exponent and are more positively asymmetric, in right- and left-handers alike, and these features depend on the functional role of the hand rather than on handedness group. These features also classify the dominant hand in unseen participants. Human movement is thus neither stereotyped nor random but complex, and multifractal analysis provides a principled way to quantify the complexity and diversity of optimized movement.

\section*{Ethics approval}
The experimental protocol was approved by the ethical committee of the Pharmacy Faculty of the Federal University of Bahia (Certificate of Presentation for Ethical Appreciation: 68289021.5.0000.5531) in accordance with the Declaration of Helsinki.

\section*{Acknowledgments}
We thank the Brazilian Federal Agency for Support and Evaluation of Graduate Education (CAPES) for supporting Yago Emanoel Ramos through grant no.\ 88887.203818/2025-00.

\appendix

\section{Hoff equation derivation}
\label{sec:appendix_hoff}

We minimize the integral of the squared jerk over a fixed duration $t_f$, with cost function
\begin{equation*}
J = \int_{0}^{t_f} \dddot{q}^2\, dt ,
\end{equation*}
and Lagrangian $L = \dddot{q}^2$. The trajectory $q(t)$ that minimizes $J$ satisfies the Euler-Poisson equation, which extends the Euler-Lagrange equation to higher-order derivatives~\cite{hoff_model_1994}:
\begin{equation*}
\frac{\partial L}{\partial q} - \frac{d}{dt}\left(\frac{\partial L}{\partial \dot{q}}\right) + \frac{d^2}{dt^2}\left(\frac{\partial L}{\partial \ddot{q}}\right) - \frac{d^3}{dt^3}\left(\frac{\partial L}{\partial \dddot{q}}\right) = 0 .
\end{equation*}
Because $L$ depends only on $\dddot{q}$, the first three terms vanish, leaving
\begin{equation*}
- \frac{d^3}{dt^3} (2\dddot{q}) = 0 \implies q^{(6)}(t) = 0 .
\end{equation*}
Integrating this sixth-order equation yields a fifth-degree polynomial for the generalized position,
\begin{equation*}
q(t) = a_0 + a_1t + a_2t^2 + a_3t^3 + a_4t^4 + a_5t^5 .
\end{equation*}
The coefficients follow from the boundary conditions of a point-to-point movement that starts and ends at rest. Let $\Delta q = q_f - q_0$ be the total generalized displacement. At $t=0$: $q(0) = q_0$, $\dot{q}(0) = 0$ and $\ddot{q}(0) = 0$, so $a_0 = q_0$, $a_1 = 0$ and $a_2 = 0$. At $t=t_f$: $q(t_f) = q_f$, $\dot{q}(t_f) = 0$ and $\ddot{q}(t_f) = 0$. Solving the remaining system gives
\begin{equation*}
a_3 = \frac{10\Delta q}{t_f^3}, \qquad a_4 = -\frac{15\Delta q}{t_f^4}, \qquad a_5 = \frac{6\Delta q}{t_f^5}.
\end{equation*}
The generalized velocity is the fourth-degree polynomial $\dot{q}(t) = 3a_3t^2 + 4a_4t^3 + 5a_5t^4$, and substituting the coefficients yields the velocity profile of Eq.~(\ref{eq:hoff}).

\section{Kinematic derivations}
\label{sec:appendix_coordinates}

Starting from the recorded positions $(x, y, z)$, the Cartesian velocity components $(v_x, v_y, v_z)$ are evaluated with second-order central finite differences with respect to the time step $\Delta t$. The scalar speed is
\begin{equation*}
|V| = \sqrt{v_x^2 + v_y^2 + v_z^2}.
\end{equation*}
To analyze the motion in different reference frames, the Cartesian state vectors are transformed into cylindrical and spherical velocity components.

\subsection*{Cylindrical velocities}
Let $\rho = \sqrt{x^2 + y^2}$ be the radial distance in the $xy$-plane. The radial velocity is $v_\rho = \dot{\rho}$, and the azimuthal (tangential) velocity $v_\phi = \rho \dot{\phi}$ is the component of the velocity perpendicular to the radial direction in the plane, equal to the $z$-component of the angular momentum per unit mass divided by $\rho$:
\begin{align*}
v_\rho &= \dot{\rho} = \frac{x v_x + y v_y}{\rho}, \\
v_\phi &= \rho \dot{\phi} = \frac{x v_y - y v_x}{\rho}.
\end{align*}

\subsection*{Spherical velocities}
Let $r = \sqrt{x^2 + y^2 + z^2}$ be the spherical radius. The radial velocity $v_r = \dot{r}$ is the projection of the velocity vector onto the position vector, and the polar velocity $v_\theta = r \dot{\theta}$ describes motion parallel to the meridians. Differentiating $\tan \theta = \rho / z$ with respect to time gives
\begin{align*}
v_r &= \dot{r} = \frac{x v_x + y v_y + z v_z}{r}, \\
v_\theta &= r \dot{\theta} = \frac{z (x v_x + y v_y) - \rho^2 v_z}{r \rho}.
\end{align*}
Together with $v_\phi$ (above), these equations give the velocities in the three coordinate systems.

\section{Brownian motion}
\label{sec:appendix_brownian}

This appendix describes the simulation of the three-dimensional confined Brownian motion shown for illustration in Fig.~\ref{fig:mfdfa}D, whose velocity components are obtained as in Appendix~\ref{sec:appendix_coordinates}.

\subsection*{Stochastic trajectory generation}
The trajectory is a discrete-time random walk of $N$ steps, corresponding to a total duration $T$ and a sampling frequency $f_s = 1/\Delta t$ (here, $T = 60$~s at $f_s = 120$~Hz, i.e.\ $N = 7200$, matching the recordings). The spatial increments in each dimension are drawn from a normal distribution,
\begin{equation*}
\Delta x, \Delta y, \Delta z \sim \mathcal{N}(0, \sigma^2).
\end{equation*}
So that the characteristic diffusion length scales with the confined volume, the standard deviation is $\sigma = \frac{L/2}{\sqrt{N}}$, where $L$ is the side length of the bounding cubic domain (here, $L = 0.5$~m, i.e.\ a volume of $0.125\,\text{m}^3$). The unconstrained positions are the cumulative sums of the increments,
\begin{equation*}
x_k = \sum_{i=1}^{k} \Delta x_i .
\end{equation*}

\subsection*{Reflective boundary conditions}
The particle is confined to the cube $[-L/2, L/2]$ in all three dimensions by a reflective boundary condition applied at each time step: a position outside the boundary is mirrored back into the domain.
\begin{equation*}
x_k \leftarrow
\begin{cases}
-L/2 + \left(-L/2 - x_k\right) & \text{if } x_k < -L/2, \\
L/2 - \left(x_k - L/2\right) & \text{if } x_k > L/2.
\end{cases}
\end{equation*}
This is applied iteratively until the position lies within the limits.

\bibliography{MM}

\end{document}